\documentclass[showpacs,amsmath,amssymb]{revtex4}
\usepackage{graphicx}
\usepackage{bm}
\begin{document}
\title{Abrupt transition in a sandpile model}

\author{Y.F. Contoyiannis}
\email{ikonto@cc.uoa.gr}
\author{F.K. Diakonos}
\email{fdiakono@cc.uoa.gr}
\affiliation{Department of Physics, University of Athens, GR-15771 Athens, 
Greece}

\date{\today}

\begin{abstract}
We present a fixed energy sandpile (FES) model which, by increasing the initial energy,
undergoes, at the level of individual configurations, a discontinuous transition.
The model is obtained by modifying the toppling procedure in the BTW \cite{BTW} rules:
the energy transfer from a toppling site takes place only to neighbouring sites with less energy 
(negative gradient constraint) and with a time ordering (asynchronous). The model 
is minimal in the sense that removing either of the two above mentioned constraints (negative gradient
or time ordering) the abrupt transition goes over to a continuous transition as in the
usual BTW case. Therefore the proposed model offers an unique possibility to explore at the 
microscopic level the basic mechanisms underlying discontinuous transitions. 
\end{abstract}

\maketitle
\section{Introduction}
Self-organized criticality (SOC) has been proposed as a universal mechanism leading to scaling laws in the 
dynamics of driven systems which evolute towards a critical state without the fine tuning of a control 
parameter \cite{BTW}. Recently it has been realized 
\cite{Grin95,HwaKar92,SorJohDor95,VespZapLor96,VesZap97,DicVesZap98} that the approach to criticality is in 
fact controled by the driving force as well as the dissipation present in the dynamics. These observations 
initiated an extensive search for the influence of the values of these control quantities to the critical 
behaviour of the corresponding SOC models. As a consequence an alternative way to study SOC models emerged 
restricting the considerations in closed systems (without external driving and dissipation) obeying the same 
dynamical rules. In these models the energy density $\rho$ is exactly conserved (FES). Since the fixed-energy
sandpiles posses simpler dynamics (without loss or addition) and are translation invariant are easier to study 
than their SOC counterparts. Furthermore the order parameter in this case can be easily identified: below a critical 
density the dynamics lead to an absorbing state characterized by the absense of activity (energy transfer processes). 
For densities above the critical $\rho_c$ the system sustains activity. Thus the critical properties of the system can 
be explored defining as order parameter the density of active sites and studying its dependence on $\rho$. In addition
the measured critical exponents in absorbing-state phase transitions can be related to avalanche exponents measured
in slowly driven systems \cite{Munoz99,Chessa98}.

Our analysis will be devoted exclusively to the sandpile models although many of our results could apply for other systems 
possessing SOC too \cite{BakSnep93}. Usually the transition from an absorbing (vanishing order parameter) to an 
active state (order parameter different from zero) in FES models is continuous \cite{Models} allowing their classification
in universality classes. On the other hand there is a variety of physical processes which are characterized by a discontinuous 
transition (melting, boiling, earthquake events, etc.). It is therefore natural to ask if such systems could be described 
in terms of fixed-energy sandpile models. In the present work we show that a suitable modification of the toppling rules in the 
BTW model can lead to a discontinuous transition. Evenmore we determine two conditions, imposed on the BTW dynamics, which are 
both neccessary and sufficient to obtain an abrupt (first order) transition in the corresponding FES model: the toppling
of an energetically activated site involves energy transfer processes only to less energetic neighbours and takes place 
sequentially in time. This observation opens the possibility to design devices with extreme sensitivity on control parameters by 
applying the analogous constraints. 

Discontinuous transitions in SOC models have already been considered in the context of the so called 
stick-slip dynamics \cite{LeuAndSor98} or the breakdown of disordered media \cite{Zapfo97}. However, our 
approach here is different as it is based entirely on slight modifications of the BTW rules. The paper
is organized as follows: in Section~2 we present the dynamics and describe the critical properties of the proposed model.
In addition we compare the obtained results with the critical behaviour of conventional FES models. In Section~3  
an intepretation (also in terms of microscopic dynamics) of the numerical results of Section~2 is given and finally in 
Section~4 we summarize our findings and give a brief outlook.
 
\section{The SMBTW FES sandpile model and its critical behaviour}

The FES model which we use is the following: we randomly depose energy on a square lattice in the form of 
grains. Each site can have an arbitrary number of grains. The total energy (and therefore the total number of 
grains) is fixed by the given energy density $\rho$. We denote as $z_{i,j}$ the number of grains
on the site $(i,j)$. A site is characterized as active if $z_{i,j}$ exceeds or is equal to a threshold value 
$z_c$. An active site topples and grains are transfered from this site ($(i,j)$) to the next neighbouring sites
provided the corresponding $z_{i \pm 1, j}$, $z_{i,j \pm 1}$ are less than $z_{i,j}$. We use the same threshold
value $z_c$ as in the original BTW model ($z_c=4$). We impose however the constraint that the toppling 
procedure is accomplished sequentially: the site $(i,j)$ can topple only if the, suitably defined, previous 
site has already toppled. The ordering of the toppling times of the lattice sites can be chosen randomly with the
constraint that the entire lattice is covered once. This sequence needs not to be the same in the following sweeps.

We have explored the dynamics of the system using a $L \times L$ square lattice and periodic boundary
conditions. As usually is the case for the FES models the control parameter is the density of grains $\rho$ 
which is a conserved quantity. As initial conditions we have used a "microcanonical" ensemble of $N$ 
configurations obtained by placing randomly, with a uniform distribution, $\rho L^2$ grains on the lattice. 
As an observable to characterize the evolution of the system within the ensemble we use the mean density of 
active sites $Q_a(m)$ at time $m$. A stationary state is described through the corresponding value of 
$Q=Q_a(\infty)$. For most of our numerical simulations we have used $N=1000$ and $L=120$. We observe that for low densities 
($\rho < 2.0495$) all initial configurations lead to the absorbing state $Q=0$. 
For $\rho > 2.0495$ (in steps $10^{-4}$ for $100 \le L \le 316$) there is a percentage $w$ of configurations which lead to $Q=0$ 
and the remaining configurations lead to $Q \approx 0.17$. In fact, in the latter case, $Q_a(m)$ approaches, in the asymptotic limit 
(for $m >> 1$), a stationary state described by random fluctuations of amplitude $0.01$ around the value 
$0.17$. In Fig.~1 we show the function $Q_a(m)$ for one such configuration at $\rho=2.0496$. For comparison we display in the same plot the 
corresponding evolution for a typical configuration with density just above the critical one ($\rho=2.1151$) in the usual BTW FES model.

\begin{figure}
\includegraphics{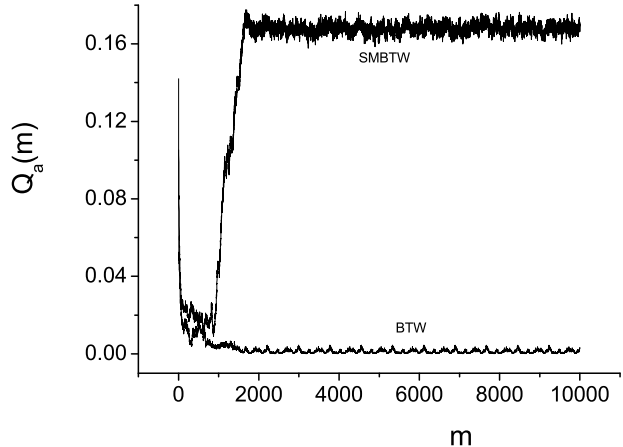}
\caption{\label{fig:fig1} The function $Q_a(m)$ for a single configuration for both the SMBTW FES model using $\rho=2.0496$ as well
as the BTW FES model ($\rho=2.1151$).}
\end{figure}

It is clearly seen that there is an energy gap between the stationary states of the system in the case of the SMBTW FES model. No such gap can
be observed in BTW FES dynamics at this scale. The value of $Q$ just above the critical density is two orders of magnitude smaller in the BTW 
than in the SMBTW case. It is then natural to assign two phases to the SMBTW system: phase A is characterized by $Q=0$ (absorbing state) 
while phase B corresponds to $Q \approx 0.17$. Then the value $\rho_c=2.0495$ represents the critical density of the system for
the given lattice size and set of initial configurations, above which the state with $Q \approx 0.17$ is accessible by the dynamics. In the neighbourhood 
of the critical value the function $Q(\rho)$ is well fitted by a sigmoidal leading to the estimation of the
critical density of the SMBTW FES $\rho_{c,SMBTW}=2.0495 \pm 0.0072$. The set of configurations leading to the 
nonvanishing value of $Q$ can be used to present the phase diagram of the model in the $(Q,\rho)$ space. 
As we show in Fig.~2a, at $\rho=\rho_{c,SMBTW}$, an abrupt jump in $Q$, possessing the caharacteristics 
of a first order phase transition, occurs. The plot $Q=Q(\rho)$ for the common BTW model is shown in Fig.~2b. As mentioned in the 
literature \cite{Bagnoli03} the nature of the phase transition in this case is not clear due to the devil's staircase form of the function 
$Q(\rho)$ \cite{Malakis01,Bagnoli03}. Our analysis however supports the scenario of a continuous transition in the 
BTW FES model. This is due to the fact that the spectrum of the stationary states $P(Q)$ accessible by the dynamics in the asymptotic limit posesses no 
energy gap in contrast to the SMBTW model. 
   
\begin{figure}
\includegraphics{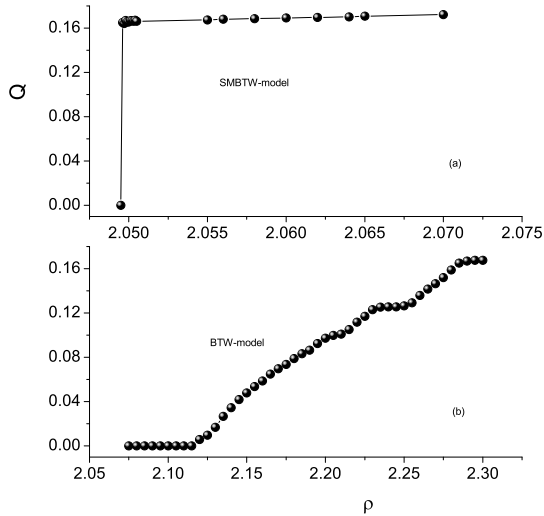}
\caption{\label{fig:fig2} (a) The order parameter $Q$ as a function of the control parameter $\rho$ of the SMBTW FES model. 
For $\rho > \rho_{c, SMBTW}$ we use, in the ensemble averaging, only configurations leading to $Q \neq 0$. (b) The corresponding plot
for the order parameter $Q$ of the FES BTW model. As in (a) we use in the ensemble averaging, for $\rho > \rho_{c,BTW}$, only configurations 
leading to $Q \neq 0$. }
\end{figure}

In order to support further our conjecture concerning the abrupt behaviour in the SMBTW FES model in contrast with a smooth transition of the usual BTW 
FES model one has to calculate the dependence of the gap of $Q$ on the lattice size $L$. We have performed a calculation of the gap in the SMBTW model for 
$5$ different lattice sizes: $80,120,200,600,1000$. For lattices with size greater than $L=120$ the value of the gap is almost constant: $Q \approx 0.169$. 
The results of this calculation are shown with crosses in Fig.~3. A solid line at $Q=0.169$ is drawn to guide the eye. Additionally we have calculated
the asymptotic value $Q$ for the BTW FES model using the same lattice sizes as for the SMBTW case. The corresponding results are presented in Fig.~3 with
full circles. It is clearly seen that the asymptotic value $Q$ in this case tends rapidly to zero with increasing lattice size. Thus, in the thermodynamic
limit, the gap in the SMBTW remains finite and large, characteristic for a first order transition, while in the BTW case no such gap occurs leading to a 
continuous transition.  

\begin{figure}
\includegraphics{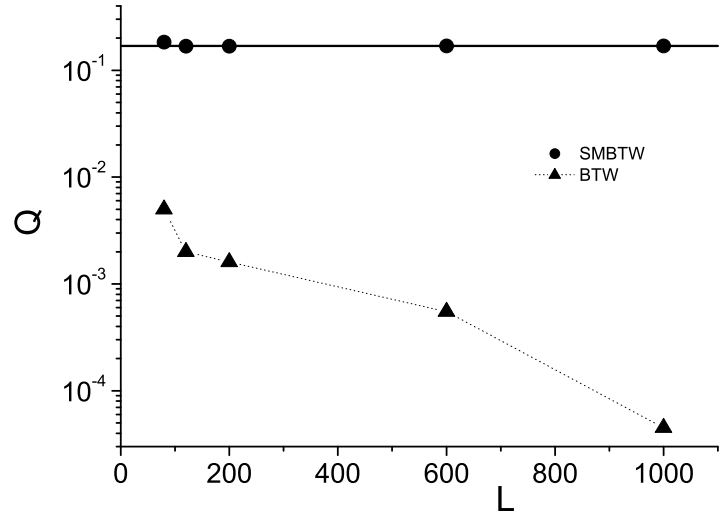}
\caption{\label{fig:fig3} The gap in $Q$ as a function of the lattice size $L$ for the SMBTW (full circles)
as well as the BTW (full triangles) FES model (semilog plot). The solid line at $Q=0.169$ is used to guide the eye.}
\end{figure}

To illustrate this property more transparently we present in Fig.~4 the distribution $P(Q)$ for the two models. To allow for a
comparison, as the critical density is different in the two cases, we calculate $P(Q)$ at densities
$\rho_i=\rho_{c,i}+ \delta \rho$ ($i=1,2$) using the same value $\delta \rho$ for the two models. 

\begin{figure}
\includegraphics{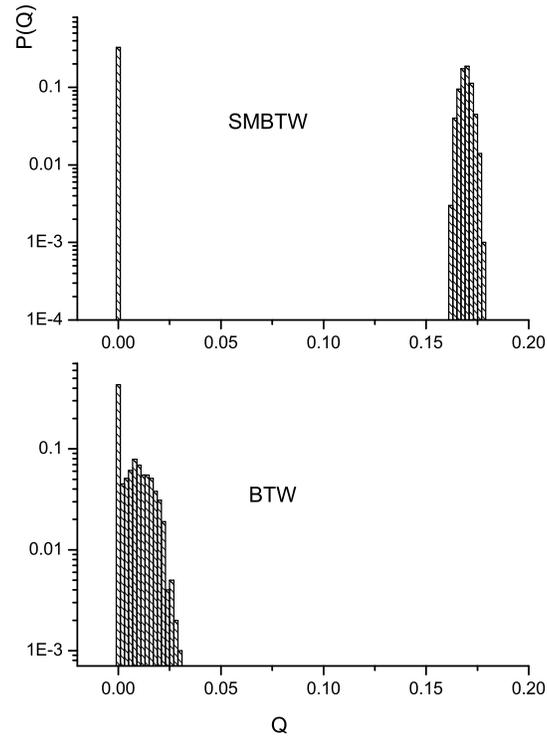}
\caption{\label{fig:fig4} The distribution $P(Q)$ for the SMBTW and BTW models at densities
$\rho_1=2.060$ and $\rho_2=2.1255$ respectively ($\delta \rho=0.0105$ for both models). }
\end{figure}

\section{Interpretation of the numerical results}
In order to understand the origin of the sharp transition in the SMBTW model we investigate the corresponding
dynamics at the microscopic level. We define the single-site states of the system in terms of the possible 
values of the occupancies (number of grains on the site) $n_s$ at a given time. To simplify our analysis
we include in the state $n_s=4$ also the (rare) case when a site is occupied with more that $4$ grains. For 
densities $\rho > \rho_c$ we observe that the single site dynamics are characterized by an ergodic behaviour: 
each one-site trajectory visits irregularly all the accesible states in phase space. After a characteristic 
time scale $t \approx 2000$ (algorithmic time in units of lattice sweeps) an invariant density is established.
Having achieved this stationarity each state is visited with almost equal probability by the dynamically 
evolving site. A very smooth maximum occurs for $n_s=1$ and $n_s=3$. As the number of possible states is $5$ 
$(n_s=0,1,2,3,4)$ we expect, for a uniform invariant density and assuming that the ergodic hypothesis applies, 
to have $1/5$ probability to be on the active state $(n_s=4)$, a value leading to $Q=0.2$ very close to the 
observed value $Q \approx 0.17$. The deviation is due to the fact that the invariant density is not exactlty 
uniform. 

It is worth exploring the global dynamics of the system as well.  Therefore we investigate the evolution
of a typical configuration of the entire lattice in the critical region $\rho \stackrel{>}{~} \rho_c$. 
In Fig.~5 we present the evolution of such a configuration for $\rho=2.060$. We show the contourplots for
the initial configuration as well as the envolving state at algorithmic time $t=5000$. We recall here that 
stationarity is achieved already at $t=2000$. To simplify our representation without loosing on physical 
insight we adopt the coarsegrained description of the phase space of the system used in 
\cite{Manna91,VesZap97} 
distinguishing between stable ($n_s < 3$, gray), critical ($n_s=3$, white) and active ($n_s > 3$, black) 
sites. For a better visualization of the details of the dynamics it is convenient to zoom into a part of the
lattice which we choose here to be the set of sites $(i,j)$ with $i,j \in [40,60]$. Obviously, as displayed 
in Fig.~5a, initially the active sites form clusters which are embedded in domains consisting of critical 
sites.

\begin{figure}
\includegraphics{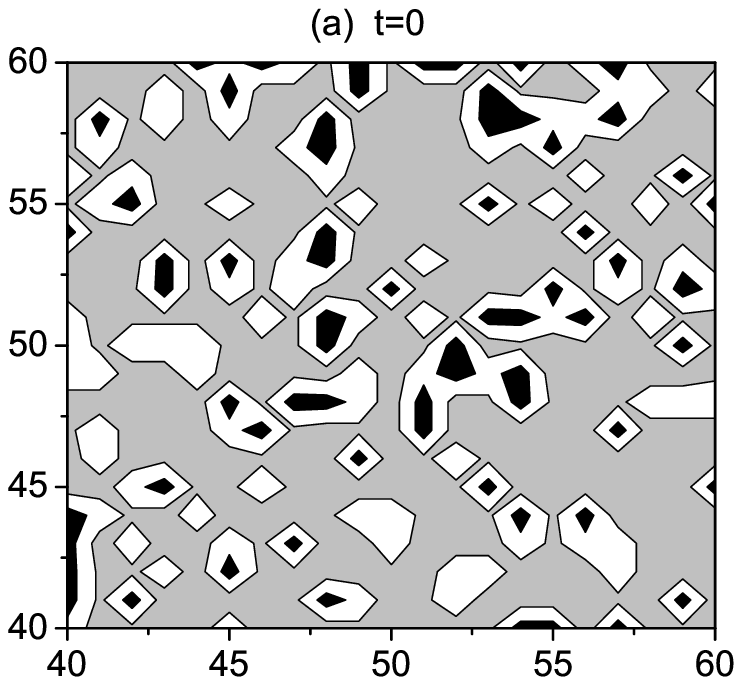}
\includegraphics{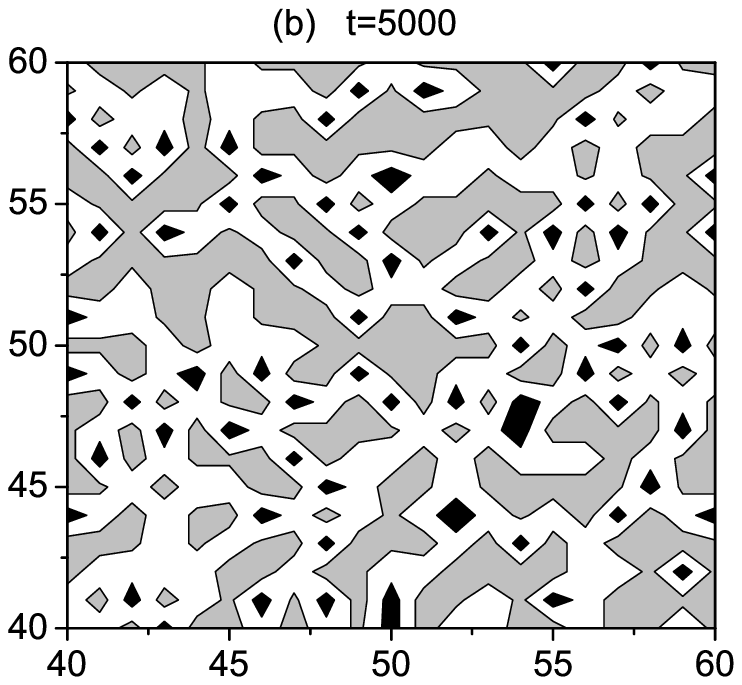}
\caption{\label{fig:fig5ab} The evolution of a typical configuration using the SMBTW rules for $\rho=2.060$.
We present the initial state (a) as well as the resulting state for $t=5000$ (b). As explained in the text 
only a part of the lattice is displayed.}
\end{figure}

As the dynamics evolve the active clusters split and their gradual destruction initiates in favour of the 
formation of larger 
domains of critical sites. Before the active clusters disappear the domains of critical sites approaching each 
other collide forming larger domains of critical sites containing smaller active clusters. When the 
stationary state is achieved the critical sites form large connected domains, similar to the 
above-threshold percolation clusters, which span the entire lattice. A dynamical equilibrium is established
characterized by an irregular deformation of the critical domains as well as a chaotic motion, icorporating
the spliting and recombining through collisions, of the active clusters within these domains. Thus the critical 
domains construct a protecting network of communicating channels (see Fig.~5b) for the irregular evolution of the active 
clusters. It must be noted that critical clusters can be formed also in the conventional sandpile models. However
the maximal size of critical clusters in the case of the FES SMBTW model is much larger than the corresponding
size of the conventional models. 

The formation of this network is a process depending very sensitively on the initial conditions. Consider
for example a typical configuration leading to $Q=0$ for $\rho=2.0495$. The total number of grains, for a 
$L=120$ lattice, in such a configuration is $25912$. Inserting one additional grain in the lattice can lead 
(depending on where we put it) to $Q \approx 0.17$. Thus, density fluctuations of the order of $10^{-5}$ may 
lead to a tremendous change of the order parameters' value. We attribute this behaviour to the first order 
character of the transition in our model. Actually imposing a time ordering in the toppling rules of the system
introduces an internal timescale of the order of the lattice size $L$. Changes in the environment of each site
during time intervals of this order are felt by the corresponding site due to the dependence of the local energy
current on the configuration of the environment expressed through the negative gradient constraint. This feedback 
mechanism creates an unstable environment which leads to a strong sensitivity to initial conditions. The relative 
timescale for which this instability influences the dynamics is of the order of $\frac{1}{L}$ and therefore 
environmental influence becomes continuous in time in the limit of an infinite system. 
We have investigated the transition by relaxing each time one of the 
two constraints we have used in the SMBTW model, i.e. the negative gradient and the time ordering. In both 
cases the transition turns out to be continuous and the corresponding phase diagram is very close to the 
diagram shown in Fig.~2b for the BTW model. Keeping only the time ordering constraint (SBTW model) the internal timescale
has no consequence on the evolution of the system as the corresponding energy currents do not depend on the
environment. Keeping only the negative gradient constraint (PMBTW model) there is no internal time and the changes
in the environment occur in timescales of the order of $L^2$ which become less and less important as the system 
size increases. On the other hand the choice of negative gradient (instead of, for example, a positive gradient) constraint 
is neccessary in order to reach a stationary absorbing state.

\section{Concluding remarks}
Let us now summarize briefly our results. In the present work we have introduced a sandpile model resulting
from the BTW rules through the addition of two constraints: energy is transfered from an active site only to
less energetic neighbouring sites and the toppling takes place in an asynchronous manner. The later means that the 
instants when a lattice site can topple are ordered in time. We have investigated the case without external driving.
The system undergoes an abrupt (first order) transition. Our model resembles the dynamics of activated
random walkers \cite{DMVZ2000} with the additional property of walking only to less occupied sites. This 
introduces a feedback mechanism influencing locally the energy flow on the lattice. The proposed model is 
minimal in the sense that both constraints are neccessary in order to achieve the discontinuous change of 
the order parameter. Considering the one-site as well as the global dynamics of the system we were able to 
understand the qualitative as well as some quantitative features of the basic mechanism underlying the observed
transition. Our model is capable to describe physical systems characterized by asynchronous spontaneous energy 
transfer. In particular the proposed model could give additional insight in prefracture processes during earthquake (EQ)
events. In this case the imposed two constraints are fulfiled: (i) the stress within the focal area is transfered from regions
of higher tension to regions of lower tension (negative gradient) while (ii) the microcrack transmission takes place 
sequentially. When the fracture of the heterogeneous environment is consumated the remaining asperities suffer an intensive 
stress from their environment. A tiny fluctuation of the surrounding stress field is decisive for the final 
fracture of asperities and therefore the occurance of an EQ event or not. In the former case the corresponding transition 
is abrupt (first order) \cite{Sornette97,CKE}. Additionally the
extreme sensitivity of the proposed model on density fluctuations of relative magnitude of the order of $10^{-5}$ 
(or even less for lattices with $L > 120$) suggests the possibility to design a high efficiency sensor by appropriate 
realization of the proposed dynamical rules. It remains a challenging task to determine in what extent the observed 
behaviour possesses universal features.

\begin{acknowledgments}
The present work was supported by the EPEAEK II (Ministry of Education) research funding programme PYTHAGORAS II.
\end{acknowledgments}

{}

\end{document}